\documentclass[twocolumn,prl]{revtex4}

\usepackage{graphicx}
\usepackage{mathptmx} 

\begin{document}

\title{
Correlations and
invariance of seismicity under
renormalization-group transformations
}
\author
{
\'Alvaro Corral
}
\affiliation{
Grup de F\'\i sica Estad\'\i stica,
Departament de F\'\i sica, Facultat de Ci\`encies, 
Universitat Aut\`onoma de Barcelona,
E-08193 Bellaterra, Barcelona, Spain 
}
\email{alvaro.corral@uab.es}
\date{\today}

\begin{abstract}
{
The effect of transformations analogous to those of the real-space
renormalization group are analyzed for the temporal occurrence
of earthquakes.
The distribution of recurrence times turns out to be invariant
under such transformations, for which the role of the correlations
between the magnitudes and the recurrence times are fundamental.
A general form for the distribution is derived imposing only the
self-similarity of the process, which also yields a scaling relation
between the Gutenberg-Richter $b-$value, the exponent characterizing
the correlations, and the recurrence-time exponent.
This approach puts the study of the structure of seismicity 
in the context of critical phenomena.
}
\end{abstract}

\maketitle


The study of statistical seismology has a long history,
exemplified by the Omori law of aftershocks and the Gutenberg-Richter
relation for the number of earthquakes above a given magnitude,
and more recently, by the fractal properties of earthquake spatial occurrence
\cite{Utsu,Kagan,Turcotte,Turcotte2}.
Less attention however has been paid to the timing of individual 
earthquakes, for which a unifying picture was missing 
until the work of Bak {\it et al.} \cite{Bak02,Corral03,Corral04b}.
The main relevance of that work was the seminal introduction
of scaling concepts in earthquake statistics, which provide 
a powerful tool to unify descriptions and to derive relations
between different quantities
(in this case, as we will see, 
by scaling we do not mean just power-law
relations, as it is sometimes assumed).
Later, Bak {\it el al.}'s procedure was modified to
study the distribution of times between consecutive events
in a single spatial region \cite{Corral04}.

Let us consider the seismicity of an arbitrary spatial region.
Given a lower bound $M_c$ for the magnitude,
and intentionally disregarding the spatial degrees of freedom,
a marked point process in time of the form
$(t_0,M_0)$, $(t_1,M_1)$, $\dots$ is obtained,
where $t_i$ denotes the time of occurrence of event $i$,
with a magnitude $M_i \ge M_c$.
The recurrence times are defined as the time intervals between 
nearest-neighbor (i.e., consecutive) events,
$\tau_i \equiv t_i-t_{i-1}$.
In the case of stationary seismicity (characterized by a long-term
linear relation between the accumulated number of earthquakes
and time),
for spatial regions of linear size
ranging from 20 km to the whole world, 
and for magnitude bounds from 1.5 to 7.5,
the probability densities $D(\tau)$ of the recurrence time
were found to verify a universal scaling law \cite{Corral04},
\begin{equation}
     D(\tau)=R f(R\tau),
\end{equation}
where $f$ is a universal scaling function and
the scaling factor $R$ is the rate of seismic occurrence, 
defined as the mean number per unit time
of events with $M \ge M_c$ in the region, 
and given by the Gutenberg-Richter law,
$R(M_c)=R_0 10^{-b M_c}$, with the $b-$value usually close to 1.


As no separation of mainshocks and aftershocks is performed,
the recurrence-time distribution consists of a mixture of different
aftershock sequences and more or less independent events;
therefore, it is very surprising that distinct regions and earthquakes of
disparate sizes present such a extreme degree of regularity.

But even more surprising,
the scaling relation for the recurrence-time distribution 
reveals that seismicity is in a highly orchestrated state,
in which the removal of events (when the lower bound $M_c$ is raised)
does not affect the properties of seismic occurrence, as the distribution
keeps the same shape (with only a different mean) independently of $M_c$.
In general, when some events are removed from a point process,
the properties of the process do change; 
therefore, the distribution of recurrence times constitutes a very
special case, invariant under a transformation akin to those
of the renormalization group (RG) in real space
\cite{Turcotte,Kadanoff,Christensen04}.

The first step of our renormalization-group transformation
consists on the raising of the lower bound $M_c$.
This implies that only a fraction of events survives the transformation,
which defines a different recurrence-time distribution.
(If $M_c$ is increased in one unit, we are dealing with an authentic
decimation, as only about 1 tenth of events are kept, 
due to the Gutenberg-Richter law.)
The second part of the procedure is the scale transformation,
which changes the time scale to make the new system comparable with the
original one.

The Poisson process, characterized by an exponential recurrence-time distribution,
represents a trivial solution to this problem when there are no correlations
in the process and therefore events are randomly removed. 
Indeed, it has been argued that the scaling function $f$ can
only be an exponential function \cite{Molchan}.
However, the scaling function clearly departs from 
an exponential \cite{Corral04}, 
and in this way the relevance of correlations in the structure of
seismicity becomes apparent.

Indeed, the scaling function $f$ can be described by a decreasing
power law for intermediate times, $R\tau < 1$, 
with an exponent about 0.3, and a faster decay for long times,
$R\tau > 1$, well approximated in both cases by a gamma distribution.
No model of earthquake occurrence is assumed to obtain
these results, they are a fundamental characteristic of seismicity.
For short times, $R\tau <0.01$, the condition of stationarity
is usually not fulfilled and the behavior is not universal.
Nevertheless, in the non-stationary case the process can be transformed 
into a stationary one with an appropriate transformation of the time
axis, and then the same scaling relation is found to hold again
\cite{Corral04}.

It should be noted that the term ``correlations'' can be understood
in two forms: If the recurrence-time distribution is not an exponential,
this implies the existence of a memory effect in the process, as events 
do not occur independently at any time, 
as it would be the case for a Poisson process.
But further, the recurrence time may depend on the history of the process,
in particular the occurrence time and magnitude of previous events.
We shall see that this type of correlations are responsible of the
breaking of the memoryless character of seismicity.

In general, the time between two consecutive earthquakes, $\tau_i$,
may depend on the magnitude of the previous event, $M_{pre}=M_{i-1}$,
the previous recurrence time, $\tau_{i-1}$, and also on the occurrence
of preceding events, $i-2$, $i-3$, etc.
In their turn, the magnitude of the $i-$th event $M_i$ can depend on $\tau_i$,
$M_{i-1}$, $\tau_{i-1}$, and so on \cite{Helmstetter}.
We shall only consider here the dependence of $\tau_i$ with $M_{i-1}$,
as we have verified it is the most important
(together perhaps with the dependence of $\tau_i$ with $\tau_{i-1}$).
Note also that although the dependence of the recurrence time and the 
magnitude with the distance between events can be important, as we have 
not considered spatial degrees of freedom, 
we do not need to take this effect into account.

So, in what follows we study the effect in the structure of seismicity
of the simplified case of correlations between
the recurrence time and the magnitude of the previous earthquake.
If we raise the magnitude threshold from $M_c$ to $M_c'$ 
the distribution of recurrence times for events with magnitudes
$M \ge M_c'$ can be obtained from the distribution 
for events with $M \ge M_c$.
Assuming that an event with magnitude $M_0 \ge M_c'$ has occurred,
we can write for the next event above (or at) $M_c'$,
\begin{equation}
\begin{array}{l}
    P[\mbox{ recurrence time $ > \tau $ for events } M \ge M_c']=\\
    P[\mbox{ 1st return time $> \tau $ and } M_1 \ge M_c' ] +\\
    P[\mbox{ 2nd return time $> \tau $ and  
             $M_1 < M_c'$ and } M_2 \ge M_c' ] +\\
    P[\mbox{ 3rd return time $> \tau $ and  
             $M_1 < M_c', M_2 < M_c'$, and } M_3 \ge M_c' ] + \cdots
\end{array}
\end{equation}
where the sum has to be extended up to infinity.
$P$ denotes probability and the $n-$th return time
is defined, for events with $M\ge M_c$,
as $t_i-t_{i-n}$, that is, as the elapsed time between any event and
the $n-$th event after it (of course, the 1st return time
is the recurrence time).
As the recurrence time depends only on the previous magnitude,
but the magnitude is independent on any other variable we can write 
\begin{equation}
\begin{array}{l}
    P[\mbox{ recurrence time $ > \tau $ for events } M \ge M_c']=\\
    P[\mbox{ 1st return time } > \tau \, | \,  M_1 \ge M_c' ] p +\\
    P[\mbox{ 2nd return time } > \tau \, | \,   
             M_1 < M_c', M_2 \ge M_c' ]
             qp +\\
    P[\mbox{ 3rd return time } > \tau \, | \,  
             M_1 < M_c', M_2 < M_c',M_3 \ge M_c']
             q^2p + \cdots
\end{array}
\label{probab2}
\end{equation}
with 
\begin{equation}
     p \equiv P[M \ge M_c' \, | \, M \ge M_c ]=10^{-b(M_c'-M_c)},
\end{equation}
where the last equality comes from the Gutenberg-Richter law
and $q\equiv 1-p=P[M < M_c' \, | \, M \ge M_c]$.

Derivation in Eq. (\ref{probab2}) with respect $\tau$ 
yields the probability densities $D$ of the different return times; 
as the recurrence times are independent on each other,
we use that the $n-$th-return-time distribution
is given by $n$ convolutions of the first-return-time distributions
(denoted by the symbol $*$) to get
\begin{equation}
   \top_{1/2} D(\tau) =
   p D_\uparrow (\tau) +qpD_\uparrow (\tau)*D_\downarrow (\tau)+
   q^2p D_\uparrow (\tau)*D_\downarrow (\tau)*D_\downarrow (\tau)
   + \cdots
\end{equation}
where 
$\top_{1/2} D(\tau)$ denotes the probability density for events
with $M \ge M_c'$ as a transformation $\top_{1/2}$ of the probability
density for events with $M\ge M_c$, $D(\tau)$;
more precisely,
$D(\tau)\equiv D(\tau \, | \, M \ge M_c)$, and
$\top_{1/2} D(\tau) \equiv D(\tau \, | \, M \ge M_c')$.
The subscript $1/2$ refers to the fact that this is only
the first half of the RG transformation.
$D_\uparrow$ and $D_\downarrow$ denote the recurrence-time
probability densities for events above $M_c$ 
conditioned to the fact that the magnitude
of the previous event is above or below $M_c'$  
($\uparrow$ or $\downarrow$), respectively. 
To be precise,
$D_\uparrow (\tau)\equiv D(\tau \, | \, M_{pre}\ge M_c', M \ge M_c)$, and
$D_\downarrow (\tau)\equiv D(\tau \, | \, M_{pre}< M_c', M \ge M_c)$.

In Laplace space the things are simpler,
\begin{equation}
   \top_{1/2} D(s) =
   p D_\uparrow (s) +qpD_\uparrow (s)D_\downarrow (s)+
   q^2p D_\uparrow (s)D_\downarrow (s)D_\downarrow (s)
   + \cdots
\end{equation}
Notice that
we have used the same symbol $D$ for both the probability 
densities and for their Laplace transforms
(which we may call generating functions), 
although they are different functions, of course.
As $q$ and $D_\downarrow(s)$ are smaller than one
(this is general for generating functions),
the infinite sum can be performed, turning out that
\begin{equation}
   \top_{1/2} D(s) =
   \frac{p D_\uparrow (s)}{1- q D_\downarrow (s)}=
   \frac{p D_\uparrow (s)}{1- D(s)+p D_\uparrow (s)},
\label{transf1}
\end{equation}
using that $D(\tau)$ is in fact a mixture of the distributions
$D_\uparrow $ and $D_\downarrow $, of the form
$D=p D_\uparrow +q D_\downarrow $.

The previous equation (\ref{transf1}) describes the first
part of the transformation. 
The second part is the scale transformation, which puts
the distributions corresponding to $M_c$ and $M_c'$ 
on the same scale. 
We will obtain this by removing the effect of the decreasing
of the rate, which has to be proportional to $p$,
so,
\begin{equation}
   \top_{1/2} D(\tau) \longrightarrow p^{-1} \top_{1/2} D( \tau /p ),
\end{equation}
and in Laplace space we get
\begin{equation}
   \top_{1/2} D(s) \longrightarrow  \top_{1/2} D( p s).
\end{equation}
Therefore, the combined effect of both transformations leads to
\begin{equation}
   \top D(s) =
   \frac{p D_\uparrow (ps)}{1- D(ps)+p D_\uparrow (ps)}.
\label{transf2}
\end{equation}

In order to get some understanding of this transformation
we can consider first the case in which there are no 
correlations between the magnitude and the subsequent
recurrence time.
Then, $D_\uparrow =D_\downarrow =D\equiv D_0$ and
\begin{equation}
   \top D_0(s) =
   \frac{p D_0 (ps)}{1- q D_0 (ps)}.
\end{equation}
The fixed points of the transformation are obtained by the solutions
of $\top D_0(s) = D_0(s) $. 
If we introduce $\omega \equiv p s$ and substitute $p=\omega/s$,
$q=1-\omega/s$,
we get, separating variables,
\begin{equation}
\frac 1 {sD_0(s)} - \frac 1 {s} =
\frac 1 {\omega D_0(\omega)} - \frac 1 {\omega} \equiv  k;
\end{equation}
where $k$ is a constant, due to the fact that 
$p$ and $s$ are independent variables 
and so are $s$ and $\omega$.
The solution is then
\begin{equation}
D_0(s)=\frac 1 {1+ ks},
\label{PoissonLaplace}
\end{equation}
which corresponds to the Laplace transform of an exponential
distribution. Indeed,
\begin{equation}
D_0(\tau)=k^{-1} e^{-\tau /k}.
\end{equation}
So, in the case in which there are no correlations in the process,
the only scale invariant distribution is, as one could have expected, 
the exponential distribution, characteristic of the Poisson process.
Let us see how the existence of correlations between the magnitudes
and the recurrence times changes this picture.

Correlations introduce new functions in the process. 
In our case, in order to iterate the transformation $\top$
we need to know how $D_\uparrow$ transforms as well.
It turns out that $D_\uparrow$ verifies an equation very similar
to Eq. (\ref{transf2}), which depends on 
$D_\Uparrow (\tau) \equiv D(\tau \, | \, M_{pre} \ge M_c'', M \ge M_c) $.
So, in order to apply again $\top$ 
we also need an equation to transform $D_\Uparrow $,
which in turn depends on higher values of the magnitude threshold.
In this way we obtain a hierarchy of equations.
An easy way to break this hierarchy is to assume that,
at least at the fixed point, $D_\uparrow $ has the same
functional form as $D$, but in a different scale.
So, let us assume
\begin{equation}
   D_\uparrow (\tau) = \Lambda D(\Lambda \tau),
\end{equation}
where $\Lambda$ depends on $M_c'-M_c$ with $\Lambda(0)=1$.

\begin{figure*}
\centering
\includegraphics[width=5.5in]{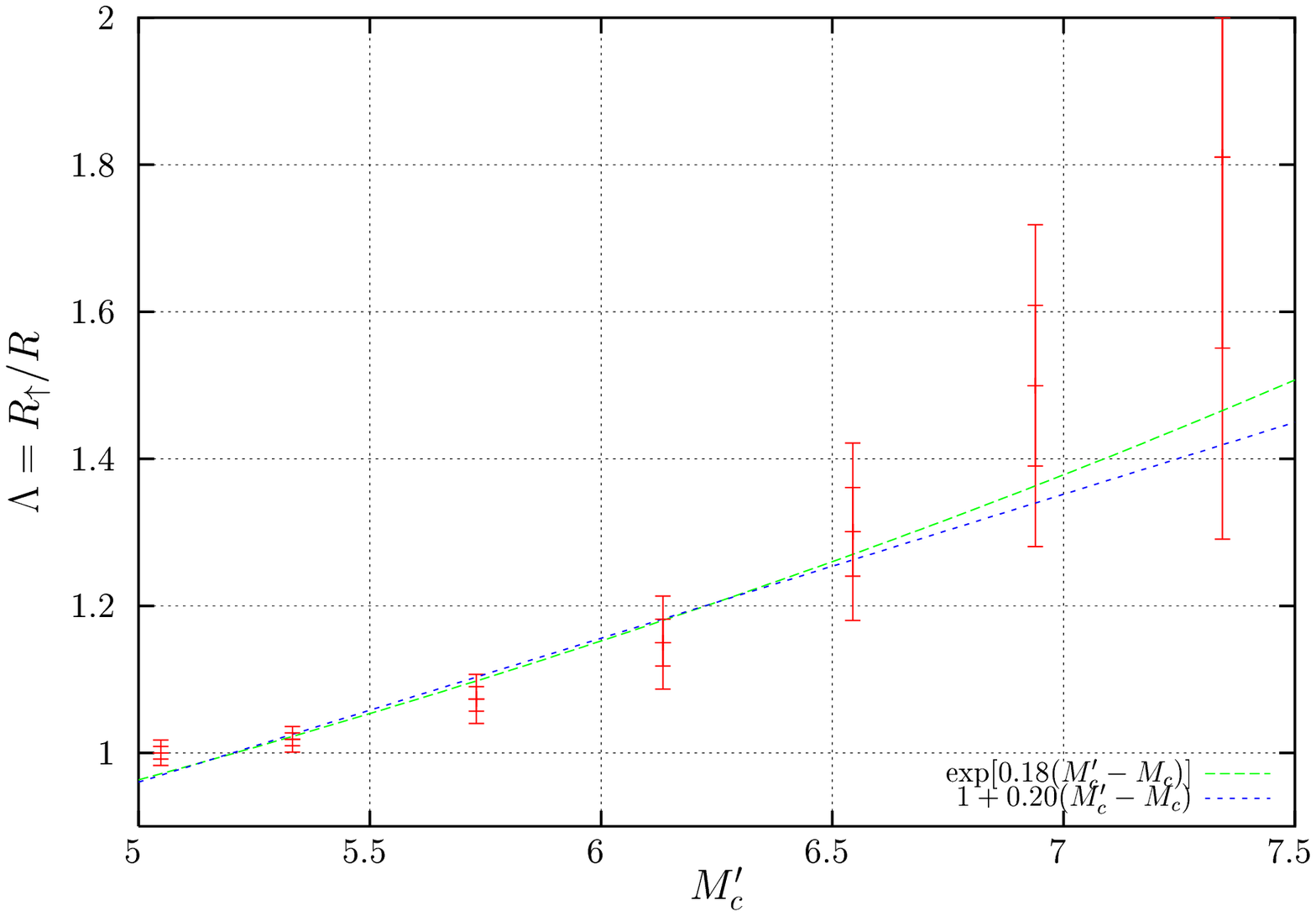}
\caption{
Inverse of the mean recurrence time, $R_\uparrow $,
scaled by $R$ for recurrence periods started by events
with $M_{pre} \ge M_c'$ and ending with $M \ge 5$.
The data correspond to worldwide earthquakes from 1973 to 2002.
The error bars mark two standard deviations of the mean
value, and the two curves are the linear and exponential fits
explained in the text. 
The last two points have not been taken into account for the fits.
\label{R_mold}
}
\end{figure*}

Figure \ref{R_mold} illustrates this behavior using worldwide
earthquakes from the NEIC PDE catalog \cite{Corral04}.
The distributions $D_\uparrow $ for different values of $M_c'$
keeping $M_c=6$ collapse onto a single curve under rescaling of the axes.
For each distribution the scaling factor is the inverse of its
mean value, $R_\uparrow $, and therefore $\Lambda=R_\uparrow /R$.
The behavior of $\Lambda$ as a function of $M_c'$ appears
in Fig. \ref{R_mold}. 
A flat line would indicate absence of correlations, as $R_\uparrow $
would be identical to $R$ and therefore $D_\uparrow =D$.
In real seismicity, $R_\uparrow $ increases with the magnitude
of the previous event, which means that the mean time between
events decreases. 
In the figure, fits of the type $\Lambda(M_c'-M_c)=A+C(M_c'-M_c)$
and $\Lambda(M_c'-M_c) = A e^{C(M_c'-M_c)}$
are shown;
in both cases it turns out to be that 
$A \simeq 1$ and $C$ is in the range $0.18 - 0.20$.

Returning to our calculation,
in Laplace space $D_\uparrow (s) = D(s/\Lambda)$;
therefore, the transformation (\ref{transf2}) turns out to be
\begin{equation}
   \top D(s) =
    \frac{p D(p s/\Lambda)}{1- D(ps)+p D(p s/\Lambda)}.
\label{transf3}
\end{equation}
%


As this discrete transformation 
is difficult to deal with, we will look at the infinitesimal transformation
defined by $M_c' \rightarrow M_c$.
Introducing $\delta\equiv M_c'-M_c$, this implies
$p=10^{-b(M_c'-M_c)}\simeq 1-B\delta$ with $B=b \ln 10$,
$D(ps)=D(s)-BsD'(s)\delta$,
$\Lambda\simeq 1+C \delta$, 
and $D(ps/\Lambda)=D(s) -(B+C)sD'(s)\delta$.
Substituting in Eq. (\ref{transf3}) and
up to first order in $\delta $ we get
\begin{equation}
   \top D(s) \simeq 
   D(s) + \{ [D(s)-1][B D(s) +C s D'(s)] - B s D'(s)\}\delta.
\end{equation}
In order to find the fixed point of the infinitesimal transformation
we impose that the coefficient of $\delta$ is zero, obtaining,
\begin{equation}
   D'(s)=-\frac{B D(s) [1-D(s)]}{s\{B+C[1-D(s)] \}},
\end{equation}
whose integration yields to 
\begin{equation}
    k s D^{1+\varepsilon}(s) + D(s)-1=0,
\label{ec2grad}
\end{equation}
where the exponent $1+\varepsilon$ comes from the definition
$\varepsilon \equiv C/B$.

We immediately see that in the case of no correlations, 
$C=0$, and therefore $\varepsilon=0$, recovering 
Eq. (\ref{PoissonLaplace}) and then 
the exponential form for $D(\tau)$.
But there are other values of $\varepsilon$ for which the previous 
equation can be easily solved.

Let us consider first the case $\varepsilon=1$, corresponding to $B=C$.
The solution of Eq. (\ref{ec2grad}) is
\begin{equation}
   D(s)=\frac{\sqrt{1+4ks}-1}{2ks},
\end{equation}
whose inverse Laplace transform can be calculated \cite{Abramowitz},
turning out to be,
\begin{equation}
   D(\tau)=\frac 1 k \left [ \frac{e^{-\tau/4k}}{\sqrt{\pi \tau /k}} +
   \frac 1 2 \left( erf\sqrt{\frac \tau {4k}} -1 \right)\right],
\end{equation}
with $erf$ the error function.
The asymptotic behavior of $D(\tau)$ is very clear:
for $\tau \rightarrow 0$ it diverges as a power law,
$D(\tau) \rightarrow 1/\sqrt{\pi k \tau} $,
whereas for $\tau \rightarrow \infty$,
$D(\tau)$ decays exponentially,
using the expansion of the error function \cite{Abramowitz}
and the fact that a power-law factor 
varies much more slowly than an exponential, 
for large arguments.

It is interesting also to study the case $\varepsilon \rightarrow 0$,
characteristic of weak correlations, $C\simeq 0$.
We can write the solution $D(s)$ as a perturbation of the
Poisson behavior corresponding to $\varepsilon=0$, i.e.,
$D(s)=D_0(s) + u(s)$, with $D_0(s)=(1+ks)^{-1}$, 
see Eq. (\ref{PoissonLaplace}).
Substituting into Eq. (\ref{ec2grad}) and using 
$(1+ks)^\varepsilon\simeq 1+\varepsilon \ln(1+ks)$ we get
\begin{equation}
    u(s)=\frac{(1+ks)^\varepsilon -1}{1+ks},
\end{equation}
and therefore $D(s)$ is just a power of the transform of the exponential
density, which means that $D(\tau)$ is a gamma distribution, i.e.,
\begin{equation}
   D(\tau)=\frac 1 {k \Gamma(1-\varepsilon)} \left( \frac k \tau\right)^
           \varepsilon e^{-\tau/k}.
\end{equation}
In this way, for $\varepsilon \simeq 0$, 
we also get a power law for short times and an
exponential decay for long times.

This behavior is by no means exclusive of $\varepsilon=1$ or
$\varepsilon \simeq 0$. 
If in Eq. (\ref{ec2grad}) we consider the limit 
$s \rightarrow \infty$ we get, as $D(s) \rightarrow 0$
(which is general for generating functions),
\begin{equation}
    D(s) \rightarrow 1/ {(ks)^\frac 1 {1+\varepsilon}},
\end{equation}
and, if $1+\varepsilon > 0$, by means of a Tauberian theorem
\cite{Feller}, 
\begin{equation}
   D(\tau) \rightarrow \frac{1}{k \Gamma[1/(1+\varepsilon)]}
   \left(\frac {k} {\tau}\right)^{\frac{\varepsilon}{1+\varepsilon}},
   \mbox{ for } \tau \rightarrow 0.
\label{shorttime}
\end{equation}
This implies that for all $\varepsilon > 0$,
which corresponds to $\Lambda > 1$,
and then to a shortening of the recurrence times
after large earthquakes, we get a decreasing power law,
which is a signature of clustering,
and in concordance with real data \cite{Corral04}.
(The situation would be reversed if $-1 < \varepsilon < 0$.)

Further insight can be obtained from Eq. (\ref{ec2grad}),
writing it as $D=1-ksD^{1+\varepsilon}$, which can be iterated
to get the Taylor expansion of $D(s)$.
Up to second order, it turns out that $D(s)=1-ks+(1+\varepsilon)k^2s^2$,
and as the coefficients of $s^n$ yield the moments of the distribution
(with a $(-1)^n/n!$ factor \cite{Feller})
we conclude that $\langle \tau \rangle = k$,
$\langle \tau^2 \rangle = 2(1+\varepsilon)k^2$
and the coefficient of variation is
$cv \equiv \sqrt{\langle \tau^2 \rangle -
\langle \tau \rangle^2}/\langle \tau \rangle = 
\sqrt{1+2\varepsilon}$.

In fact, the Taylor expansion can be obtained up to any order,
and so, all the moments $\langle \tau^n \rangle$ exist and
$D(\tau)$ decays faster than any power law for $\tau \rightarrow \infty$,
in agreement again with the observations \cite{Corral04}.

To conclude, we note that both $B$ and $C$ can be understood as critical
exponents. 
As the energy radiated by an earthquake is about $E=E_0 10^{3M/2}$ 
\cite{Turcotte},
the Gutenberg-Richter law writes $P[E \ge E_c] \propto 1/E_c^{2b/3}$.
Also, assuming the exponential form for $\Lambda$, 
$ R_\uparrow=\Lambda R= R  e^{C(M_c'-M_c)}$ and we get
\begin{equation}
   \langle \tau \rangle_\uparrow \equiv R_\uparrow ^{-1} 
   \propto \left(\frac 1 {E_c'}\right)^{\frac{2C}{3 \ln 10}}.
\end{equation}
If we define $\nu = 2C/(3\ln 10)$, this exponent, the $b-$value, and
the exponent of the recurrence-time
distribution for short times given by Eq. (\ref{shorttime})
fulfill the following scaling relation,
\begin{equation}
    \frac \varepsilon {1+\varepsilon}=\frac C {B+C} =\frac \nu {\nu + 2b/3},
\end{equation}
just using $\varepsilon=C/B$ and $B=b \ln 10$. 
In fact, we must understand this relation as the contribution of the
$M_{pre}-\tau$ correlations to the recurrence-time distribution,
as the value obtained for $C$ (about 0.2) is too small to account
for the value of the exponent, about 0.3, using a $b-$value $\simeq 1$.
We believe the inclusion of other correlations in the calculation 
will yield to a better quantitative agreement.

Our approach mainly consists of a simplification of real seismicity,
which allows to understand the complex structure of seismic occurrence
in the time and magnitude domains.
It is remarkable that simply by imposing the self-similarity of the process
and with the only assumption of the scaling of the 
conditional distributions $D_\uparrow $
(which is in agreement with the observations),
we get such a general characterization of the recurrence-time distribution.
With this study we have shown how 
the structure of seismic occurrence in time, space, and magnitude
can be understood as a critical phenomenon and then
constitutes a statistical-physics problem \cite{Bak96}.

The Ram\'on y Cajal program and the projects
BFM2003-06033  (MCyT) and
2001-SGR-00186 (DGRGC) are acknowledged.

\newpage

\begin{figure*}
\centering
\includegraphics[width=5.5in]{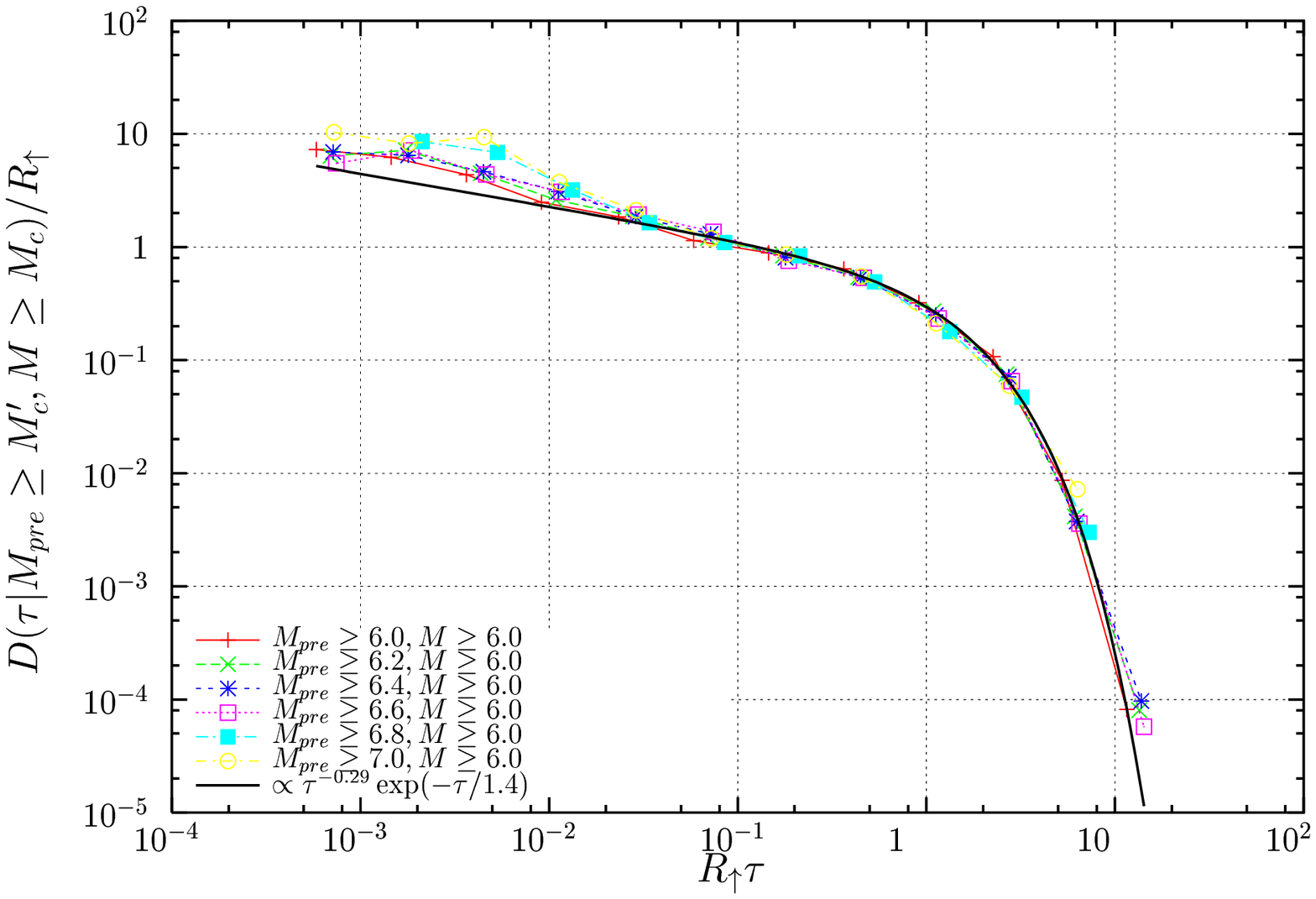}
\caption{
Recurrence-time probability densities 
conditioned to $M_{pre} \ge M_c'$,
$M_c'$ varying from 6 to 7 and with $M \ge 6$,
for worldwide earthquakes from 1973 to 2002.
The data collapse indicates that $D_\uparrow (\tau)$ verifies
the same scaling relation as $D(\tau)$. The straight line is a fit
to $D(\tau)$, turning out to be $\propto e^{-\tau/1.4}/\tau^{0.29}$.
\label{Dtcond}
}
\end{figure*}

\end{document}